\documentclass[referee,A4paper]{raa}            

\usepackage{graphicx,times}             
\begin{document}

   \title{Multi-wavelength Fibril Dynamics and Oscillations Above Sunspot - I. Morphological Signature}

   \volnopage{Vol.0 (200x) No.0, 000--000}      
   \setcounter{page}{1}          

   \author{E. S. Mumpuni
      \inst{1,2}
   \and D. Herdiwijaya
      \inst{1}
   \and M. Djamal
      \inst{3}
   \and T. Djamaluddin
      \inst{2}
   }

   \institute{Graduate Program in Astronomy, Faculty of
Mathematics and Natural Sciences, Bandung Institute of Technology, Indonesia; {\it nggieng@students.itb.ac.id}\\
        \and
             Space Science Center, LAPAN, Indonesia\\
        \and
             Graduate Program in Physics, Faculty of
Mathematics and Natural Sciences, Bandung Institute of Technology, Indonesia\\
   }

   \date{Received~~2009 month day; accepted~~2009~~month day}

\abstract{In this work we selected one particular fibril from a high resolution solar chromosphere observation from the Dutch Open Telescope, and tried to obtain a broad picture of the  intricate mechanism that might be incorporated in the multiple layer of the Solar atmosphere in high cadence multi-wavelength observation. We analyzed the changingvfibril patter using multi-wavelength tomography, which consists of both H$\alpha$ line center \& the blue wing, Doppler-signal, Ca II H, and the G-band. We have found that the intermittent ejected material through fibril from Doppler images has clearly shown oscillation mode, as seen in the H$\alpha$ blue wing. The oscillations in the umbrae and penumbrae magnetic field lines that are above the sunspot cause a broadening and forms the area like a ring shape from 3 to 15-minute oscillations as function of height. These made a distinct boundary of umbrae and penumbrae which suggest the comb structure, and indicate that the oscillations could propagate along the inclined magnetic flux tubes from below. The 3-minute strongly appeared in the broadly inclined penumbrae magnetic filed lines and gave the clear light-bridge. The well known 5-minute was dominated in the umbrae-penumbrae region boundary, the long 7-minute one was transparent in the H$\alpha$ blue wing, but this was the same with 10 and 15-minute, it was concentrated in the inner-penumbrae, as seen in the H$\alpha$ line center. From these findings we propose a picture on the role of fibril as the fabric of interaction between the layers, also the related activites around the active region under investigation.
}
   \authorrunning{Mumpuni, Herdiwijaya, Djamal \& Djamaluddin}            
   \titlerunning{Multi-wavelength Fibril Dynamics - I. Morphological Signature}  

   \maketitle

%
%
\section{Introduction}           
\label{sect:intro}

The intricate coupling interactions among photosphere-chromosphere-corona have not known yet. The structure of photosphere is relatively well understood but not so for chromosphere, in particular the dynamic of energy and mass transport in fibril, as is much discussed in Rutten (\cite{2012RSPTA.370.3129R}). Instead of mass transfer, i.e. downflow and upflow from photosphere to chromosphere and corona also exhibits a variety of oscillation modes, such as 3-minute, 5-minute, 7-minute, etc. They act as one physical package that resemble the mass and energy transfer from photopheric to chromospheric layers, and vice-versa.

Recently, high spatial and temporal resolution sequence images by ground and space based observations confirmed and revealed the number of oscillation modes (Nagashima et al.~(\cite{2007PASJ...59S.631N}), Reznikova \& Sibasaki~(\cite{2012ApJ...756...35R}), Jess et al.~(\cite{2012ApJ...757..160J})). Naturally, the disturbance or instability of magnetic structures will trigger the oscillations. Observations of sunspot oscillations were known for almost four decades  (Beckers 
\& Tallant~(\cite{1969SoPh....7..351B}), Bogdan 
\& Judge~(\cite{2006RSPTA.364..313B}) and references therein). The 5-minute oscillations (Leighton et al.~\cite{1962ApJ...135..474L}) that predominantly occur above the sunspot in the photosphere have been relatively well studied. Their amplitudes decrease with height, and they can hardly be detected in the upper chromosphere and transition region. Lites~(\cite{1992sto..work..261L}) summarized the 3-minute oscillations and drew the possible conclusion that the resonance response by many oscillation modes from below and fast mode wave driving.

Moreover, Centeno et al.~(\cite{2006ApJ...640.1153C}) also suggested for propagating shock waves in umbrae regions. The 3-minute oscillations could be due to a complex interaction of many processes, especially magneto-acoustic mode conversion and intrinsic reduced acoustic emissivity in strong magnetic fields (Stangalini et 
al.~\cite{2012A&A...539L...4S}). So, it was clear that the oscillations and wave propagation are the key processes for carrying energy and mass through differents atmospheric layers, before dissipating them into other modes.

By analyzing the oscillations and wave propagation we can derive physical mechanism about the stratiﬁcation and dynamics of different magnetized atmospheric structures. The conversion into thermal heating, e.g. coronal heating becomes the big question and the most challenging research in solar physics. The magnetic field lines are mostly believed to have the crucial role. However, there were still many unanswered problems related to these oscillation modes and their connection with other dynamic structures like fibrils and light bridge. 

This work will study the chromospheric dynamic and how it related to the photosphere by studying the sunspot umbrae, penumbrae and fibril from high resolution multiwavelength image sequence observations. The results from the Dutch Open Telescope (DOT) were analysed as diagnostic tool for tomography of the Sun (Rutten et 
al.~\cite{2004A&A...413.1183R}). The selection of H$\alpha$ images as the primary tool present a much more complete proxy to delineate chromospheric magnetic topology which consists of the mass flow along fibrils (Rutten,~\cite{2007ASPC..368...27R}). In this study we found morphological signature that indicating mass motion which may related to wave propagation

To study how the dynamic on each particular layer of the atmosphere, we use the tomographic image analysis method for each particular features. Such as the G-band, Ca II H line and H$\alpha$ which represent the activity in the photosphere, the lower chromosphere, and upper chromosphere, respectively. We will also address several oscillation modes of umbrae and penumbrae by simultaneously using time series multi-wavelength observations. We believe that this powerful technique most stringently detects wave propagation. 

In Section~\ref{S-D&A}, we briefly describe the DOT observations and the data reduction procedures. Some interesting results on the fibril mass transfer and oscillations are reported in Section~\ref{S-Res}. Discussions on the fibril and light bridge oscillations are given in Section~\ref{S-Disc}.

\section{Data and Analysis}
\label{S-D&A}

\begin{figure}    
   \centerline{\includegraphics[width=0.5\textwidth,clip=]{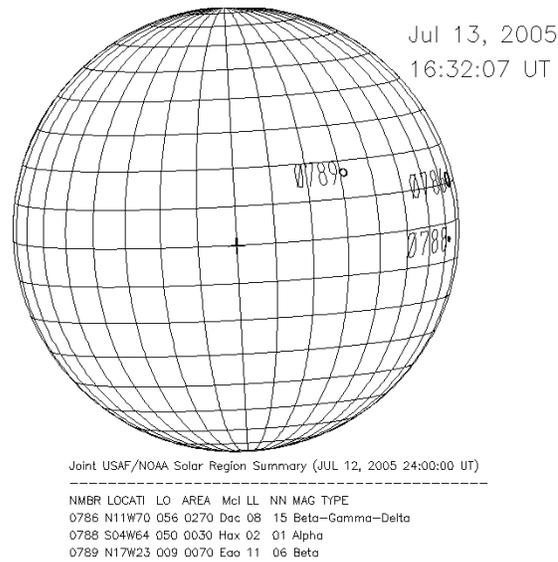}
              }
   \caption{The position of Active Region 10789. The region has McIntosh classification Eao, magnetic type $\beta$.}
    \label{20050713}
  \end{figure}

The sunspot of NOAA Active Region 10789 (N17W23) that is located close to the disk center on July 13, 2005, was selected for this study (Figure~\ref{20050713}). This active region had a simple bipolar with beta and Eao for magnetic and McIntosh classifications, respectively. There were no flares reported within three days before and after the date of observation. A series of high angular resolution images of a sunspot at AR 10789 on July 13, 2005 have been carefully filtered and aligned from the observation results of Dutch Open Telescope. The multiple wavelengths of Ca II H line, G-band, H$\alpha$ line center, H$\alpha$ blue-wing and H$\alpha$ Doppler-signal images were taken simultaneously on that time. 

DOT capabilities on multi-wavelength re-imaging system, the instrumentations, speckle acquisition and reconstruction already discussed by Rutten et al.~\cite{2004A&A...413.1183R}.

\begin{figure}    
   \centerline{\includegraphics[width=0.7\textwidth,clip=]{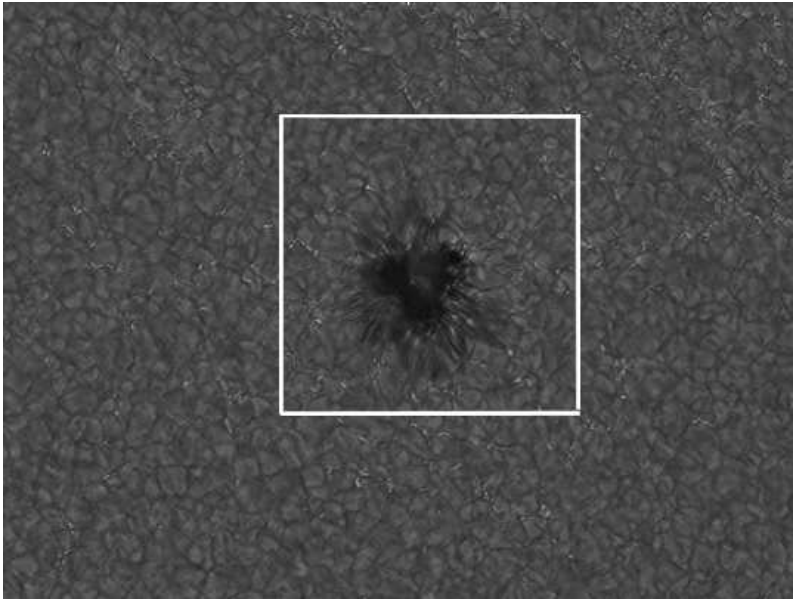}
              }
   \caption{NOAA Active Region 10789 on 2005 July 13, from the G-band images. White box is the selected area under study from multi-wavelength analysis. The picture size is 76".25 $\times$ 57".79 with  resolution 0.071''/pix.}
    \label{selected_region}
  \end{figure}

The data has already been speckle reconstructed from observational run on 07:49 - 10:35 UT to give resolution of 0.071"/pix. There are 332 images in total with a time cadence of 30 seconds for each band. The interesting feature of a light bridge appeared across the fairly round main umbrae, instead of long fibril across the penumbra (Figure~\ref{selected_region}). To see the dynamics of the region we selected a particular area of interest and studied the Doppler-signal images and made the image difference from Doppler-signals  which will show a particular feature of fibrilar pattern of our interest. Then, we manually determined a clear and persistent feature for the region under investigation, as shown in Figure~\ref{selected_region}. 

\section{Results}
\label{S-Res}

\subsection{Fibril Mass Transfer}

\begin{figure}    
   \centerline{\includegraphics[width=1.0\textwidth,clip=]{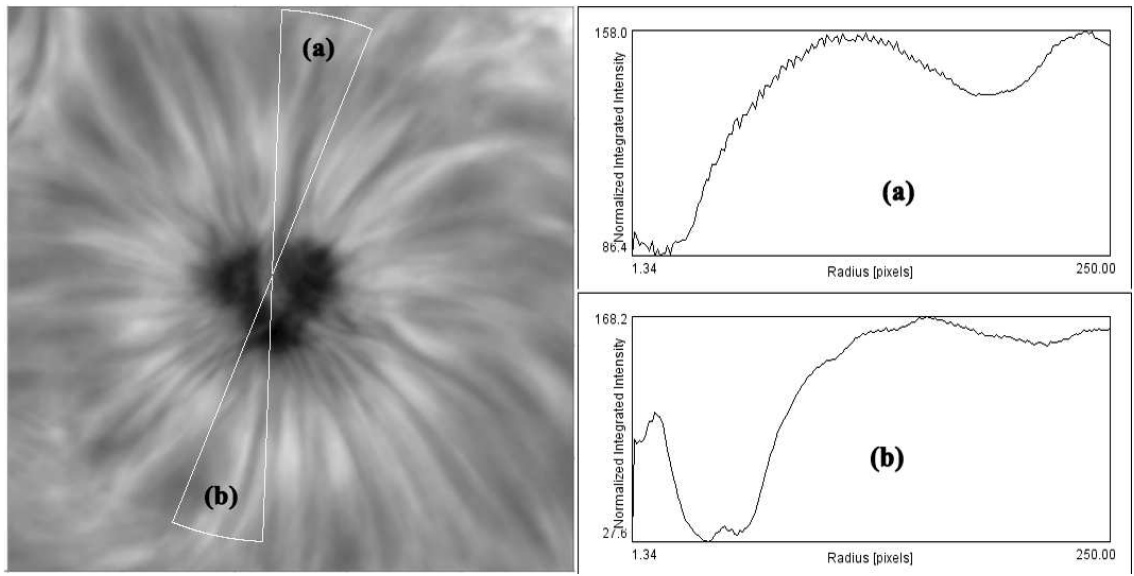}
              }
   \caption{In the H$\alpha$ line center image, the fibril under investigation is marked in the upper yellow circular sector (a), while the opposite lower circular sector is selected as comparison (b). The graphs show the integrated cross-section of the area under investigation, top panel for fibril area (a) and the bottom panel for the opposite direction (b).}
    \label{fibril_profile_all_rev2}
  \end{figure}

Figure~\ref{fibril_profile_all_rev2} illustrates the fibril under investigation (region (a), upper circular sector) from H$\alpha$ line center, and lower circular sector as comparison (region (b)). The angle of the sector is 10 degree. The right panels show the integrated cross-section of the upper (a) and lower (b) circular sector. The radius of the umbrae regions has 52 pixels and the penumbrae region extends to 105 pixels. The selection of circular sector is based on assumption that magnetic field will stronger inside the umbrae region compared to the outside region. Using this method, we can deduce that there was fibrilar pattern under region (a), which has about 150 pixels radius in length. The normalized integration intensity (total integrated intensity is divided by the number of pixels in the circle) of region (a) was higher in the umbrae area than in region (b), but about the same in the penumbra area. So, we will make further analysis of region (a).

\begin{figure}    
  \centerline{\includegraphics[width=0.7\textwidth,clip=]{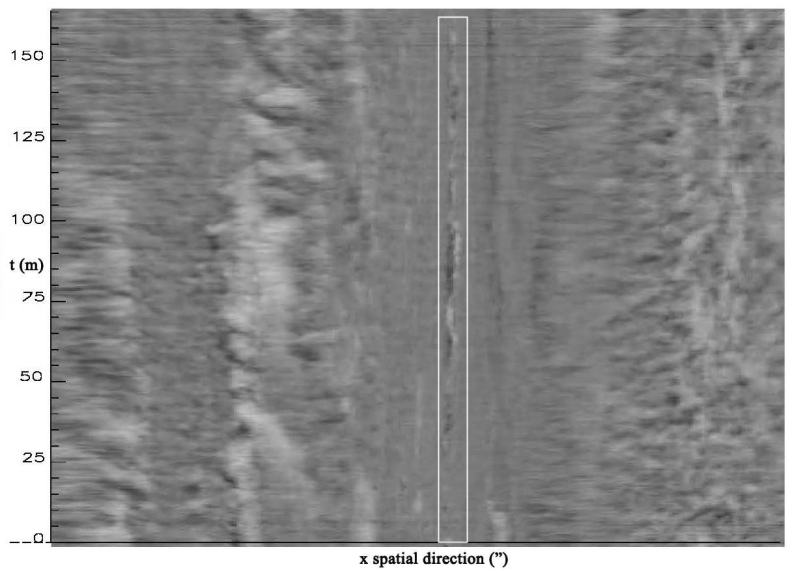}
              }
   \caption{Time slice in Doppler-signal image of selected region where the x-axis represents the x direction and the ordinate represents the time slice of 332 images, and the area under investigation is in the white rectangle. The center of the rectangle is the middle of circular sector (a) on Figure~\ref{fibril_profile_all_rev2}, with the direction of spatial direction is along the x direction, and below to top is propagation in time.}
    \label{slice_ar_rev2}
  \end{figure}

After the area selection in region (a), we then follow how the chosen fibril pattern changes overtime by slicing the central part of the selected region, to see how the mass transfer changed over the sequential times. We argue that the pattern of mass transfer from this morphological signature related to filamentary structures with filling mass proposed by Rutten,~\cite{2006ASPC..354..276R}. The same procedures were conducted for other wavelengths for H$\alpha$ blue wing, H$\alpha$ line center, Ca II H, and G-band. From the H$\alpha$ Doppler-signal, it appeared that there was a pattern of alternating features, right in the middle of the picture, marked with the white rectangle with 7 pixels in width, as shown in Figure~\ref{slice_ar_rev2}. The results of area profile for each band were shown in Figure~\ref{slice_ar_multi_cross_section}. Each time marker is equal to 30 seconds.

\begin{figure}    
   \centerline{\includegraphics[width=0.75\textwidth,clip=]{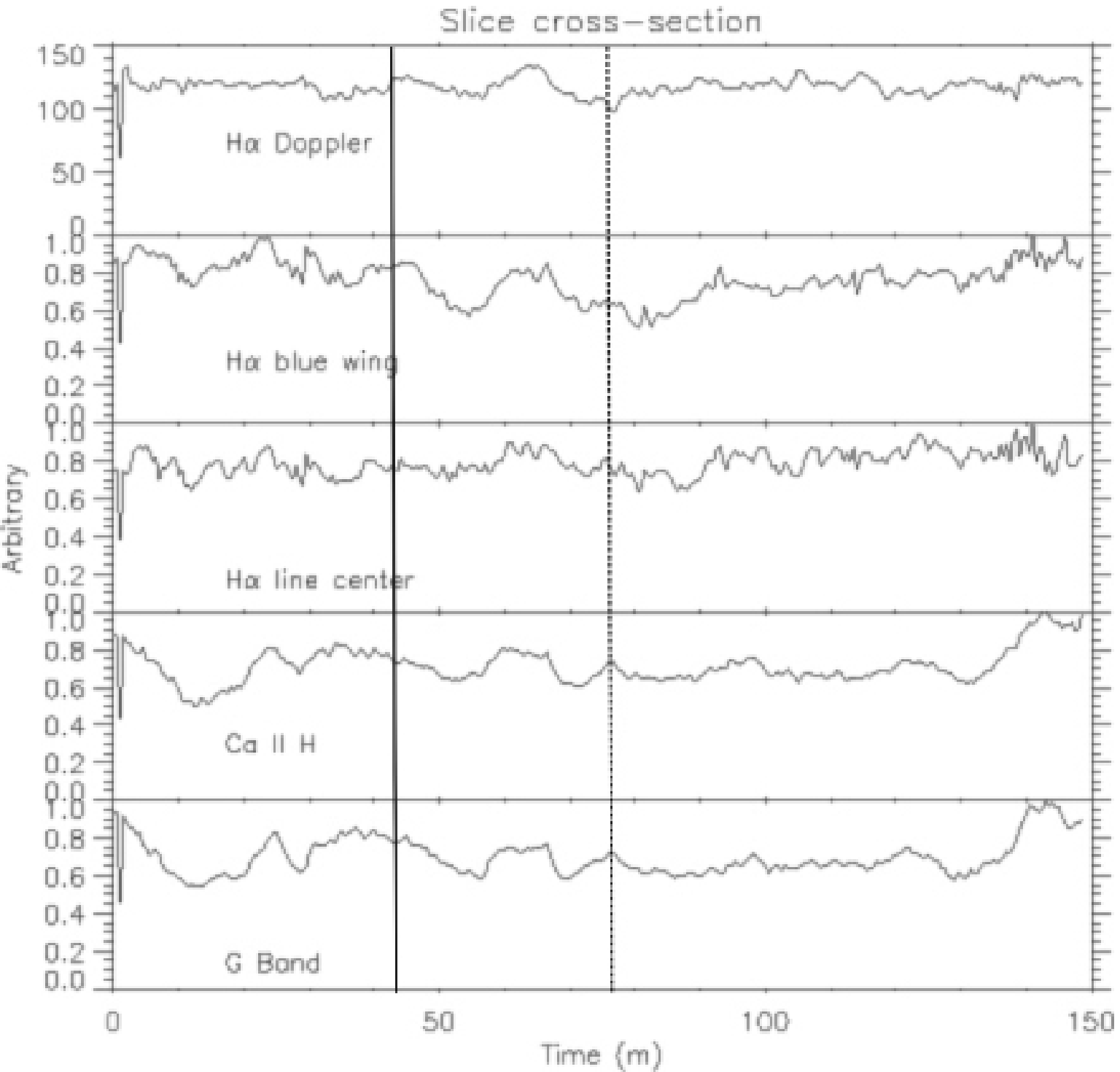}
              }
   \caption{Cross-section of the time slice of various multi-wavelength images, top to bottom: H$\alpha$ Doppler-signal, blue-wing and line center, Ca II H and G-band. The x axis is the time direction of the slice and y as arbitrary intensity  of 332 images.}
    \label{slice_ar_multi_cross_section_rev2}
  \end{figure}

We follow the time propagation in detail from the whole slice data. The upflow mass transfer was clearly seen from photosphere to chromosphere layers, see the line on the left in Figure~\ref{slice_ar_multi_cross_section_rev2}. However, the downflow was also occurred from the chromosphere, as seen by dashed line on the right.

In fibril, we find that upward motion predominantly occurred from below and alligned with the magnetic field, as seen in G-band and Ca II H, respectively. During that short period, the material packed from below loaded to the upper or lower layer and change the pattern of oscillation in the H$\alpha$ Doppler-signal and H$\alpha$ blue wing. It is seen that the mass transfer flow is not a continuous pattern and has rapid changes of less than 30 seconds, especially in the upper level which suggested that geometrically fibril is a tree-like structure.

\subsection{Umbrae, Penumbrae and Fibril Oscillations} 
\begin{figure}    
   \centerline{\includegraphics[width=1.0\textwidth,clip=]{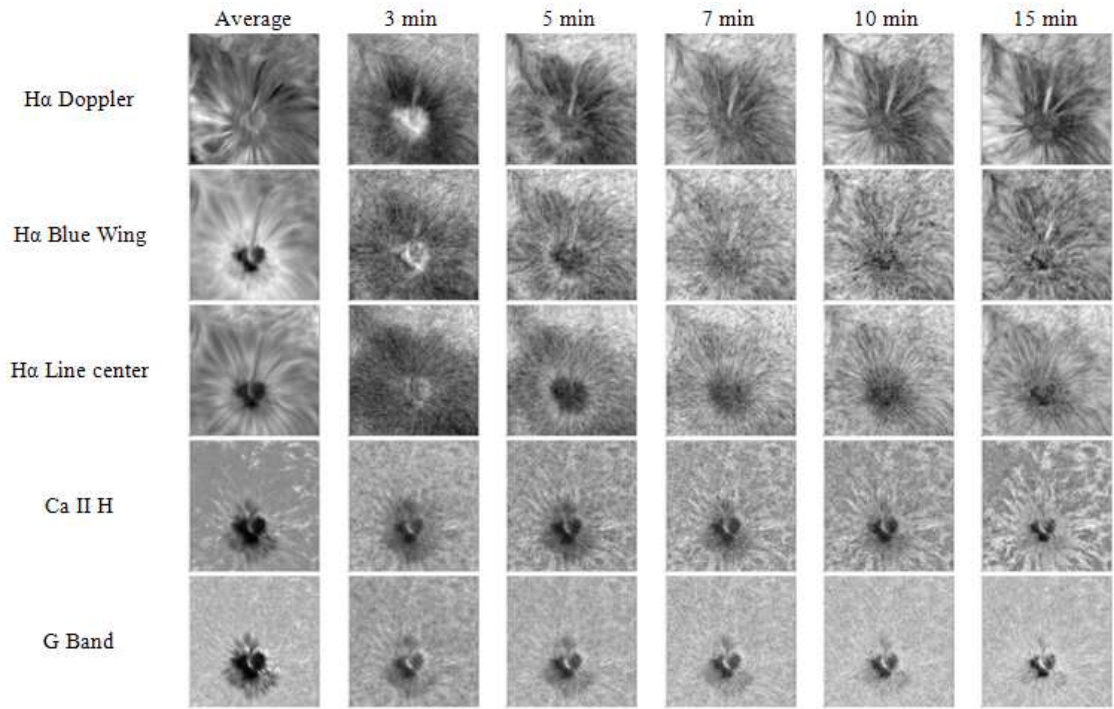}
              }
   \caption{Comparison of Average (1st column) with power-maps on 3,5,7,10 and 15-minute (second to sixth column respectively), on several wavelength: H$\alpha$ Doppler (1st row), H$\alpha$ blue wing, center line, Ca II H and G-band (second to fifth row respectively). The size of window is 512$\times$512 pixels.}
    \label{powermap}
  \end{figure}

In order to estimate the strength of an oscillation, the squared amplitude, or power, and in our case as the “power-map”, was computed from the Fourier transform (White 
\& Athay,~\cite{1979ApJS...39..317W}). Then we created 2-D power-maps of several periods on multi-wavelength bands (G-band, Ca II H$\alpha$, H$\alpha$ line center and H$\alpha$ blue wing) at 3, 5, 7, 10 and 15-minute, and the H$\alpha$ Doppler-signal as shown in Figure~\ref{powermap}. McAteer et al.~(\cite{2002ApJ...567L.165M}) shown that several spectral signatures detected near the 7, 10 and 15-minutes. The method for power-map discussed in Krijger et 
al.~(\cite{2001A&A...379.1052K}). H$\alpha$ Blue Wing is the best proxy-magnetometer to locate and spatially track extended magnetic elements, this line is the "chromospheric" line that can be used as photospheric diagnostic (Leenaarts et 
al.~\cite{2006A&A...452L..15L}). This work will address relations of oscillation in umbrae, penumbrae and fibril which are important to study the relation between the fibril and the sunspot (Chae et al.~\cite{2014ApJ...789..108C}).

The Doppler-signal indicates that there was an oscillation pattern that appeared during several moments in between material loading, and to understand how the materials flows along fibrils, we need to compare oscillatory signatures, both radial and along the fibril.

To see the layer by layer dynamics of the atmosphere, we use the power-maps from the multi-wavelength. The 3-minute which mainly observed in the chromosphere, 5-minute predominantly photospheric, and the longer period of sunspot oscillations that were given by Staude~(\cite{1999ASPC..184..113S}) and Bogdan~(\cite{2000SoPh..192..373B}). McAteer et al.~(\cite{2002ApJ...567L.165M}) suggested the presence of multiple peaks in the power spectrum with periods in the 4-15-minute range.

For each region (a) in Figure~\ref{powermap}, we follow the pattern of fibril under investigation, as indicated in Figure~\ref{fibril_profile_all_rev2}. From visual inspection in Figure~\ref{powermap}, we found that the fibrilar pattern appear strongly in the H$\alpha$ blue wing, in particular with reversal pattern (bright) in 15-minute oscillations (6$^{th}$ column, 2$^{nd}$  row). 

However, for the larger picture of the active region, we found that for this H$\alpha$ blue wing, there is also a pattern of brightening in the umbra region, in the 3-minute oscillations (2$^{nd}$ column, 2$^{nd}$ row). The same brightening pattern also appeared in the H$\alpha$ line center (2$^{nd}$ column, 3$^{rd}$ row) and H$\alpha$ Doppler-signal (2$^{nd}$ column, 1$^{st}$  row) in Figure~\ref{powermap}.

It was seen that the oscillations above sunspot in the umbrae show a dark and bright ring shapes as a function with height and distinctively divide umbrae and penumbrae which suggests the comb structures, and indicate that the oscillation could propagate along the inclined magnetic flux tubes from below.

\begin{figure}    
   \centerline{\includegraphics[width=0.6\textwidth,clip=]{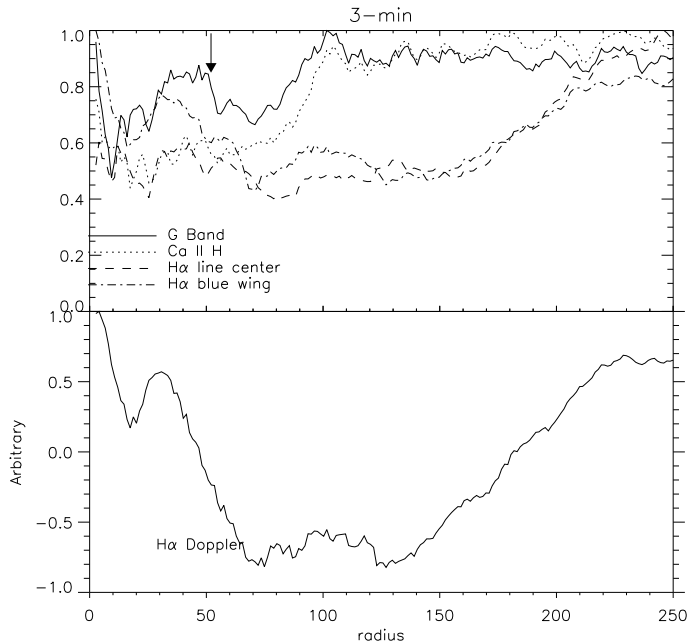}
              }
   \caption{Integrated profile for 3-minute mode of the circular sector of the fibril in region (a) as described in Figure~\ref{fibril_profile_all_rev2} for each wavelength, with Doppler-signal comparison in lower panel. The arrow sign marks the umbrae area for the smaller radius, and penumbrae region on the larger radius.}
    \label{3m}
  \end{figure}

\begin{figure}    
   \centerline{\includegraphics[width=0.6\textwidth,clip=]{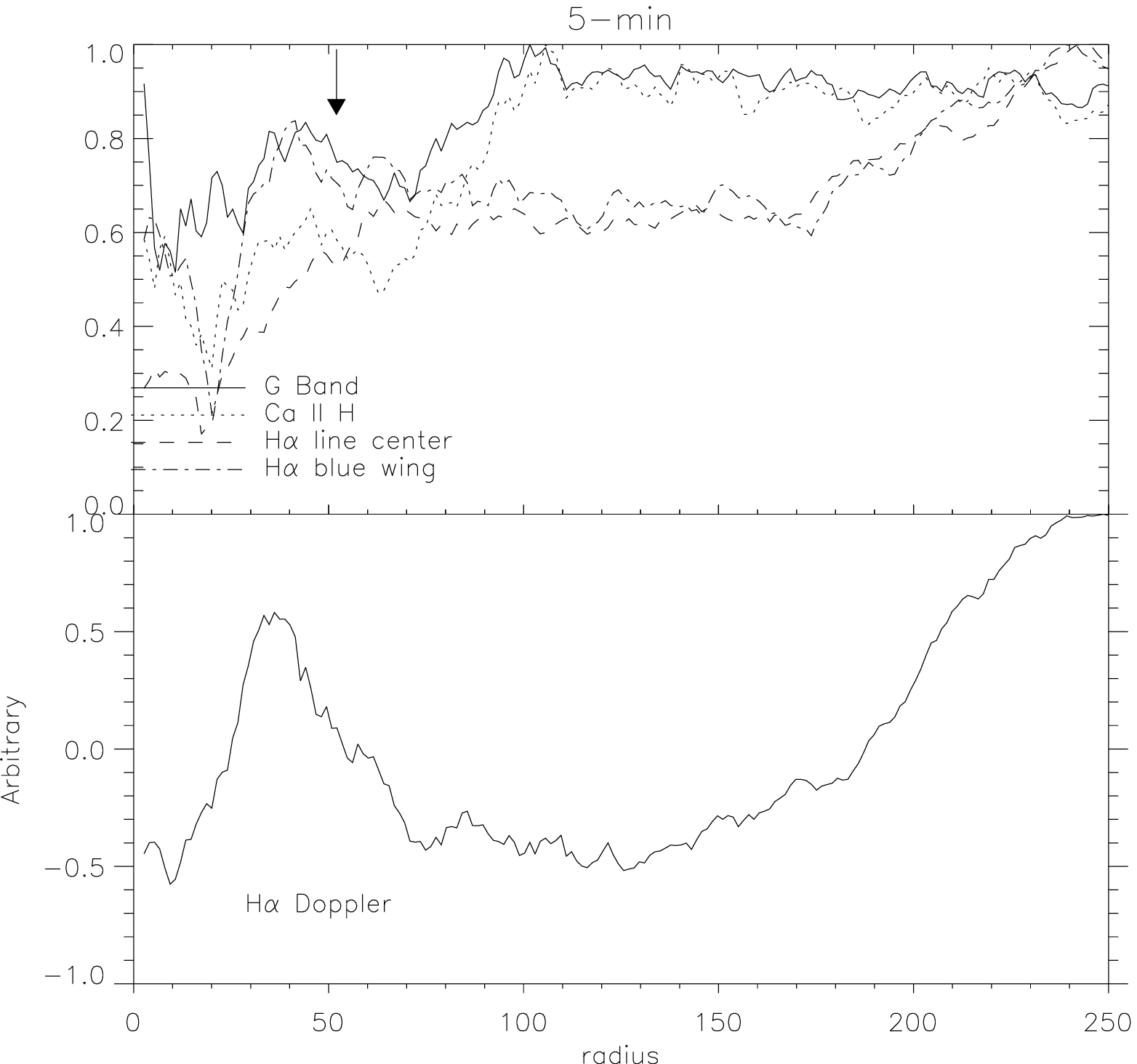}
              }
   \caption{Integrated profile for 5-minute mode of the circular sector of the fibril in region (a) as described in Figure~\ref{fibril_profile_all_rev2} for each wavelength, with Doppler-signal comparison in lower panel. The arrow sign marks the umbrae area for the smaller radius, and penumbrae region on the larger radius.}
    \label{5m}
  \end{figure}

\begin{figure}    
   \centerline{\includegraphics[width=0.6\textwidth,clip=]{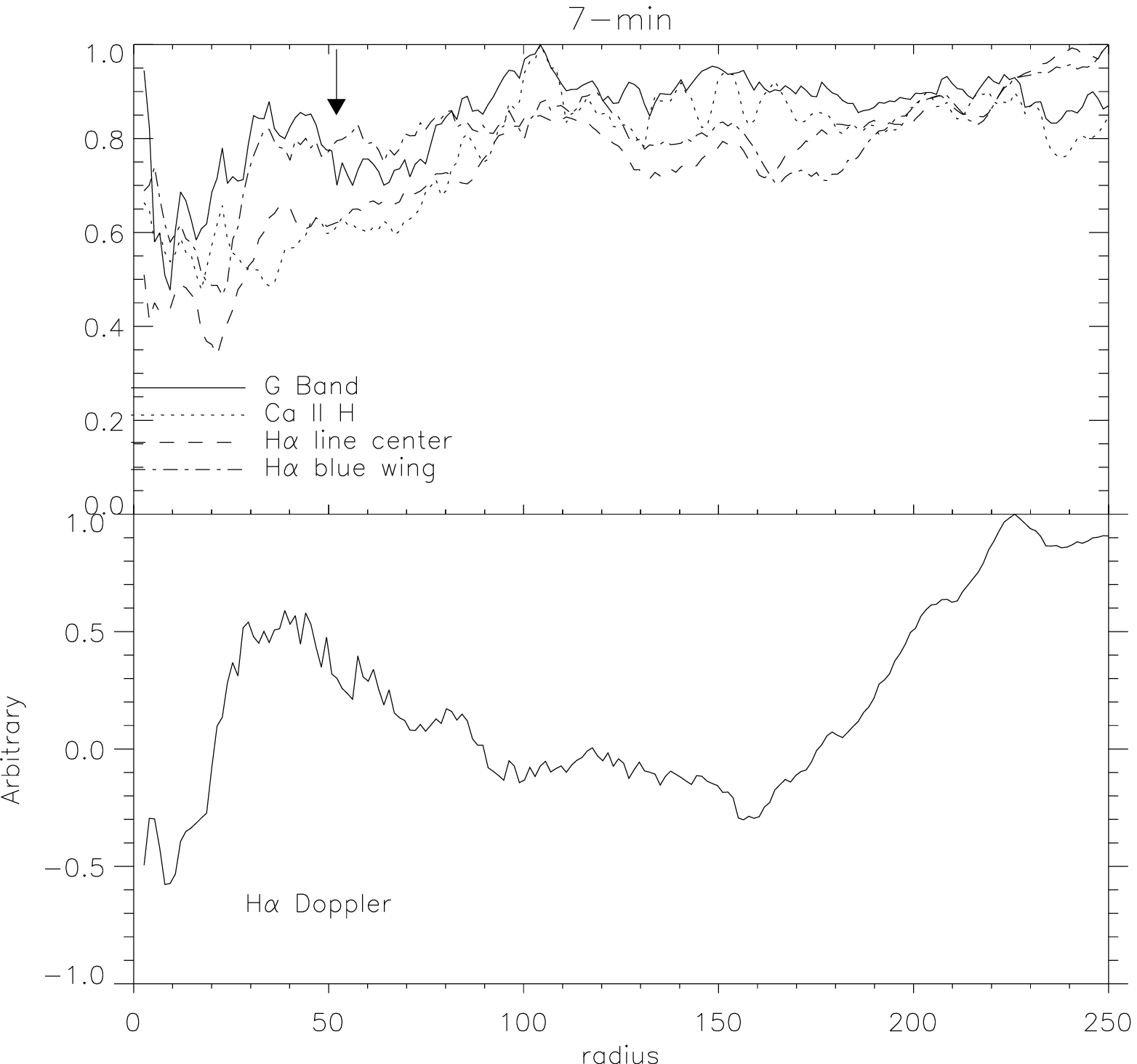}
              }
   \caption{Integrated profile for 7-minute mode of the circular sector of the fibril in region (a) as described in Figure~\ref{fibril_profile_all_rev2} for each wavelength, with Doppler-signal comparison in lower panel. The arrow sign marks the umbrae area for the smaller radius, and penumbrae region on the larger radius.}
    \label{7m}
  \end{figure}

\begin{figure}    
   \centerline{\includegraphics[width=0.6\textwidth,clip=]{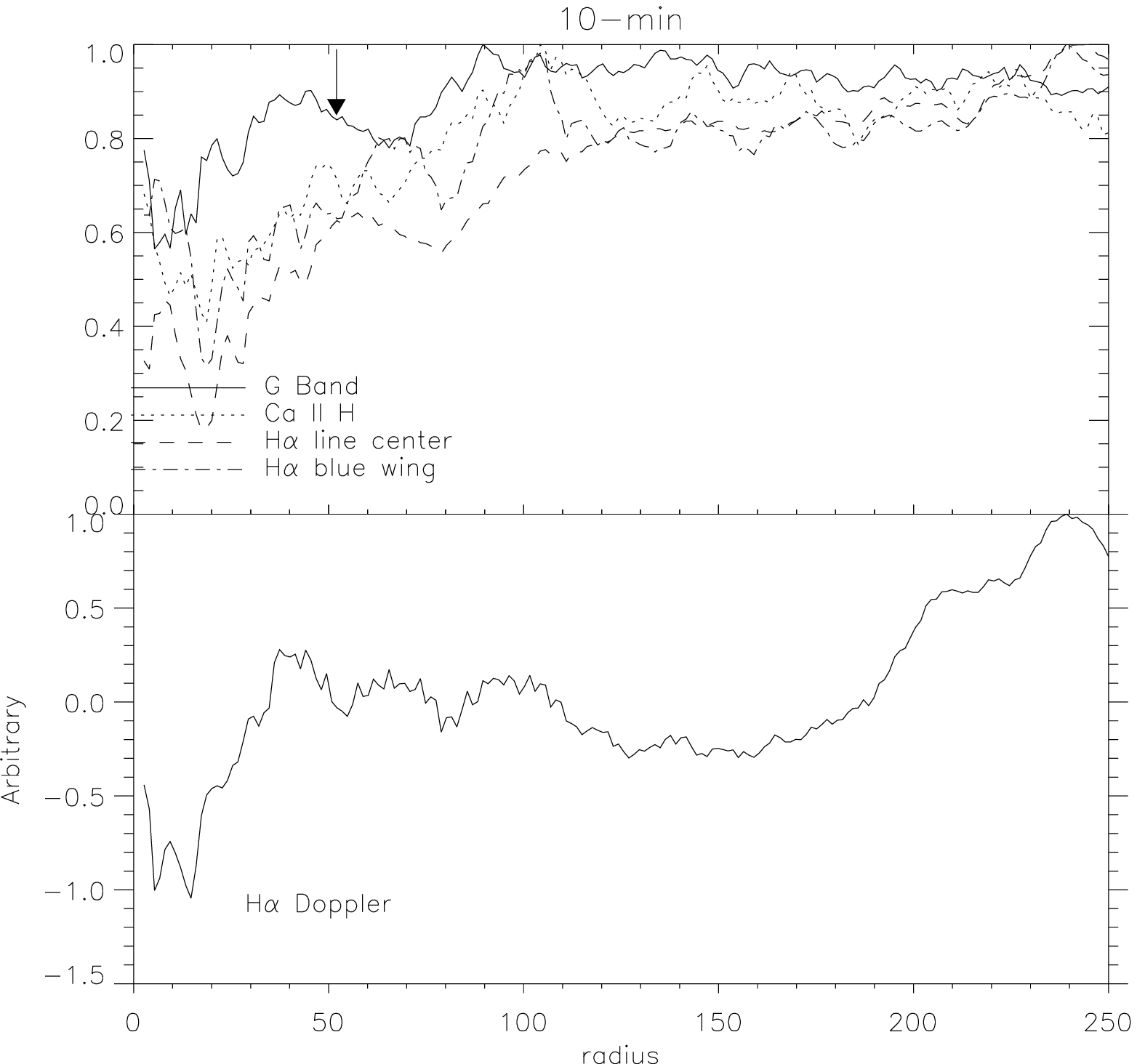}
              }
   \caption{Integrated profile for 10-minute mode of the circular sector of the fibril in region (a) as described in Figure~\ref{fibril_profile_all_rev2} for each wavelength, with Doppler-signal comparison in lower panel. The arrow sign marks the umbrae area for the smaller radius, and penumbrae region on the larger radius.}
    \label{10m}
  \end{figure}

\begin{figure}    
   \centerline{\includegraphics[width=0.6\textwidth,clip=]{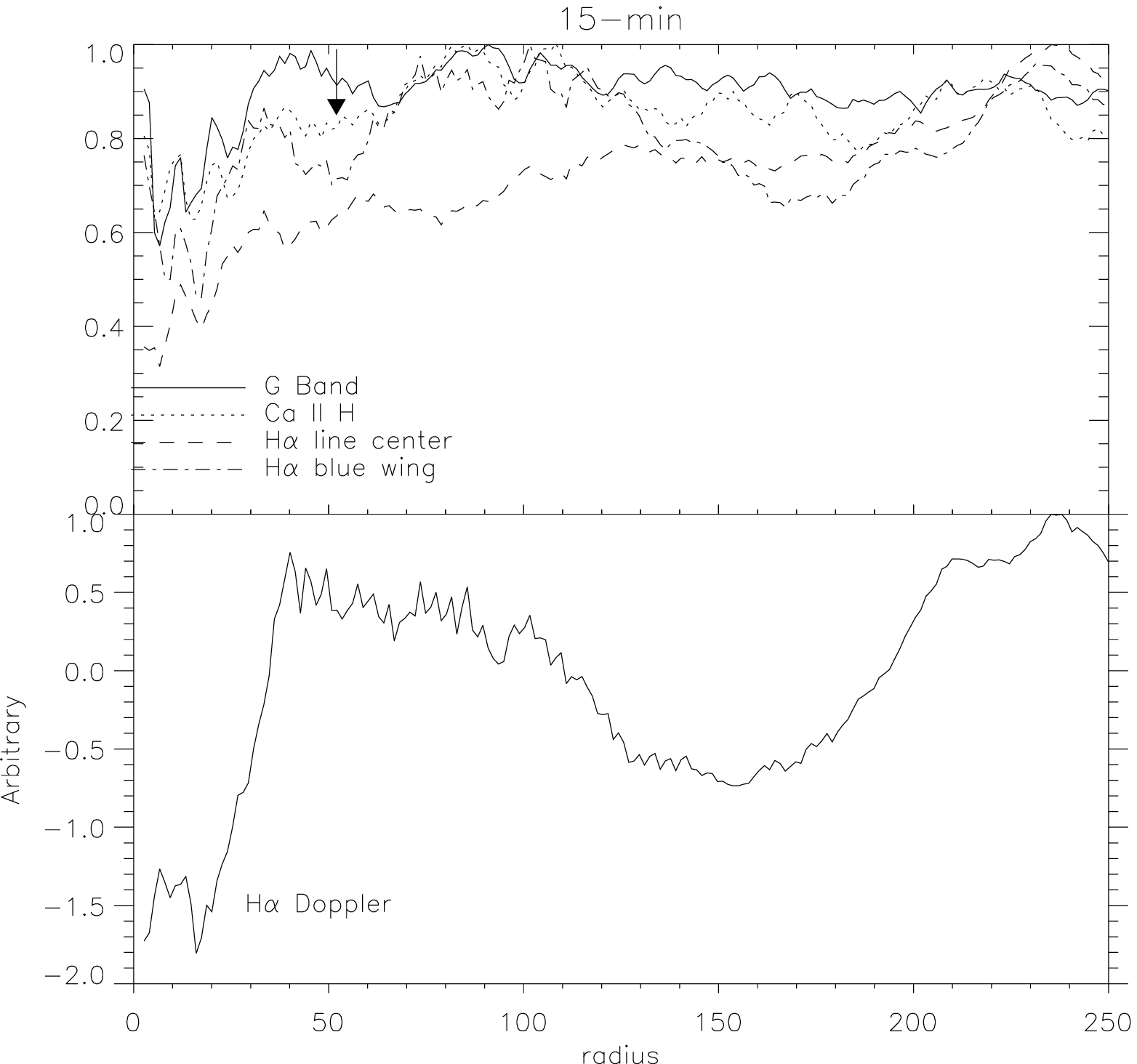}
              }
   \caption{Integrated profile for 15-minute mode of the circular sector of the fibril in region (a) as described in Figure~\ref{fibril_profile_all_rev2} for each wavelength, with Doppler-signal comparison in lower panel. The arrow sign marks the umbrae area for the smaller radius, and penumbrae region on the larger radius.}
    \label{15m}
  \end{figure}

Figure~\ref{3m} to Figure~\ref{15m} shows the radius profile of the fibril in region (a) for each mode of power for each wavelength, with Doppler-signal as \add{a} comparison on the lower panel for each mode. The arrow in each figure from Figure~\ref{3m} to Figure~\ref{15m}  marks the separation between umbrae and penumbrae.

\subsubsection{3-minute Oscillations}
In Figure~\ref{powermap}, the 3-minute oscillation strongly appeared in the bright area of H$\alpha$ blue wing due to the broad inclined penumbrae magnetic filed lines in which magneto-acoustic waves could be channeled along them. It is seemed that the 3-minute oscillation could also have broader area from photosphere to the chromosphere, even through the solar coronal fan (Jess et al.~\cite{2012ApJ...757..160J}).

Moreover, the 3-minute oscillations give the clear boundary of the light bridge and ejected material, as clearly seen in the H$\alpha$ line center and H$\alpha$ blue wing. The brightening in the umbrae region, as seen in Figure~\ref{powermap} correspond to the downflow motion as shown in its profile in Figure~\ref{3m}. The downflow patterns happened in the inner and outer bright ring. The penumbral and fibril region showed upflow and downflow patterns, respectively. The 3-minute responses were distinctly different in photosphere and chromosphere, particularly in the penumbrae region which may be due to the magnetic field strength and its inclination (Schad et al.~\cite{2013ApJ...768..111S}). 

Ca II H line has an agreement with G-band in the penumbrae region. We showed that the dynamic in the boundary of umbrae-penumbrae shows more complex picture and is crucial for physical mechanisms. From the H$\alpha$ Doppler-signal, the inner umbrae flow shows the opposite direction in 3-minute contra 5-minute to 15-minute, which indicates that 3-minute oscillations might play an important role in downward mass motion  for the fibril. There was a strong enhancement on the  Doppler-signal in the 3-minute oscillation that might cause the enhancement of the chromospheric line of H$\alpha$ and might lead to umbral flash (Rouppe van der Voort et 
al.~\cite{2003A&A...403..277R}).

In the quiet region, i.e. witha a radius more than 190 pixels, there was no difference among indicator lines for 3-minute oscillation in which gradually reverse to a downward motion.

\subsubsection{5-Minute Oscillations}
The well known 5-minute (Leighton et al.~\cite{1962ApJ...135..474L}) was dominated in the outer umbrae region near the boundary of umbrae-penumbrae, as in the H$\alpha$ blue wing, which suggests the comb structure that predominantly the downward motion, see Figure~\ref{5m}. It was related to the downward motion, in contrast with the inner umbrae. In the boundary of umbrae and penumbrea, there was a so-called "terminal region" in Doppler shift which represents the same response of photospheric and chromospheric indicator lines. The characteristics in the penumbrae and quiet region are the same as 3-minute oscillations. Both oscillations modes were sensitive to magnetic field strength and its inclination, as well as mass motion flow. 

\subsubsection{7-Minute Oscillations}
In Figure~\ref{7m}, the 7-minute has the same pattern in the H$\alpha$ blue wing and G-band in all areas which means that this oscillation was not too sensitive in the photosphere, chromosphere, penumbrae and quiet area. The difference in the umbrae magnetic field and inclination may cause this kind of oscillation. Some downflow motion was still occurred in the umbrae region. The downflow was more dominant in the larger terminal region of the radius, but the inner penumbrae showed an upward mass transfer. Moreover, the penumbrae region did not show significant downward flow.

\subsubsection{10-Minute Oscillations}
In Figure~\ref{10m}, there was no large fluctuation in flow along the near boundary of umbrae-penumbrae to the boundary of the  penumbrae-quiet area. All chromospheric lines had the same pattern. This suggests that 10-minute excites in the chromospheric layer with a relatively small magnetic field strength and inclination.

\subsubsection{15-Minute Oscillations}
In Figure~\ref{15m}, there was an upward motion in the inner umbrae and the boundary of the penumbrae-quiet region which influence the intensity of this oscillation. H$\alpha$ line center was damped compared to the Ca II H. Except in the inner umbrae, all lines intensity was relatively the same value. As for 7-minute oscillations, this mode was not significantly influenced by magnetic field changes.

\subsection{Root of Fibril}
\begin{figure}    
   \centerline{\includegraphics[width=0.4\textwidth,clip=]{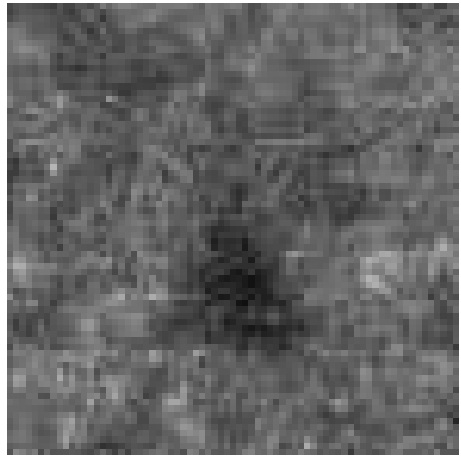}
              }
   \caption{The power-map of 15 minutes 1600 $\AA$, resized on spatial resolution with other powermap pictures of Figure~\ref{powermap}.}
    \label{powermap_1600}
  \end{figure}

We tried to compare the result with TRACE data (Handy et al.~\cite{1999SoPh..187..229H}), particularly the 1600 $\AA$, where Krijger et 
al.~(\cite{2001A&A...379.1052K}) argue that the radiation from the "temperature minimum region", based on the work of Vernazza et al.~(\cite{1981ApJS...45..635V}). There were 18 data-sets available for the window of the same observation, with 9 minutes cadences of 768 $\times$ 768 pixels for 1600 $\AA$, 0.5"/pixel. We made power-spectrum with 15-minute for the 1600 $\AA$, as shown in Figure~\ref{powermap_1600} and manually alligned the data with other data to have similar region of interest.

From Figure~\ref{powermap_1600}, it can be seen that there was no significant feature that appeared in the 15-minute 1600 \AA, except the inner umbrae. Since the “temperature minimum region” formed below, near the photosphere, at h = 500 km above continuum optical depth $\tau_5$ = 1 at $\lambda$ = 5000 $\AA$ and defines the transition from photosphere to chromosphere (Krijger et 
al.~(\cite{2001A&A...379.1052K}), Vernazza et al.~(\cite{1981ApJS...45..635V})), it appeared that the root of the fibril formed somewhere else.


\section{Discussion}
\label{S-Disc}

    \begin{figure}    
  \centerline{\includegraphics[width=1.0\textwidth,clip=]{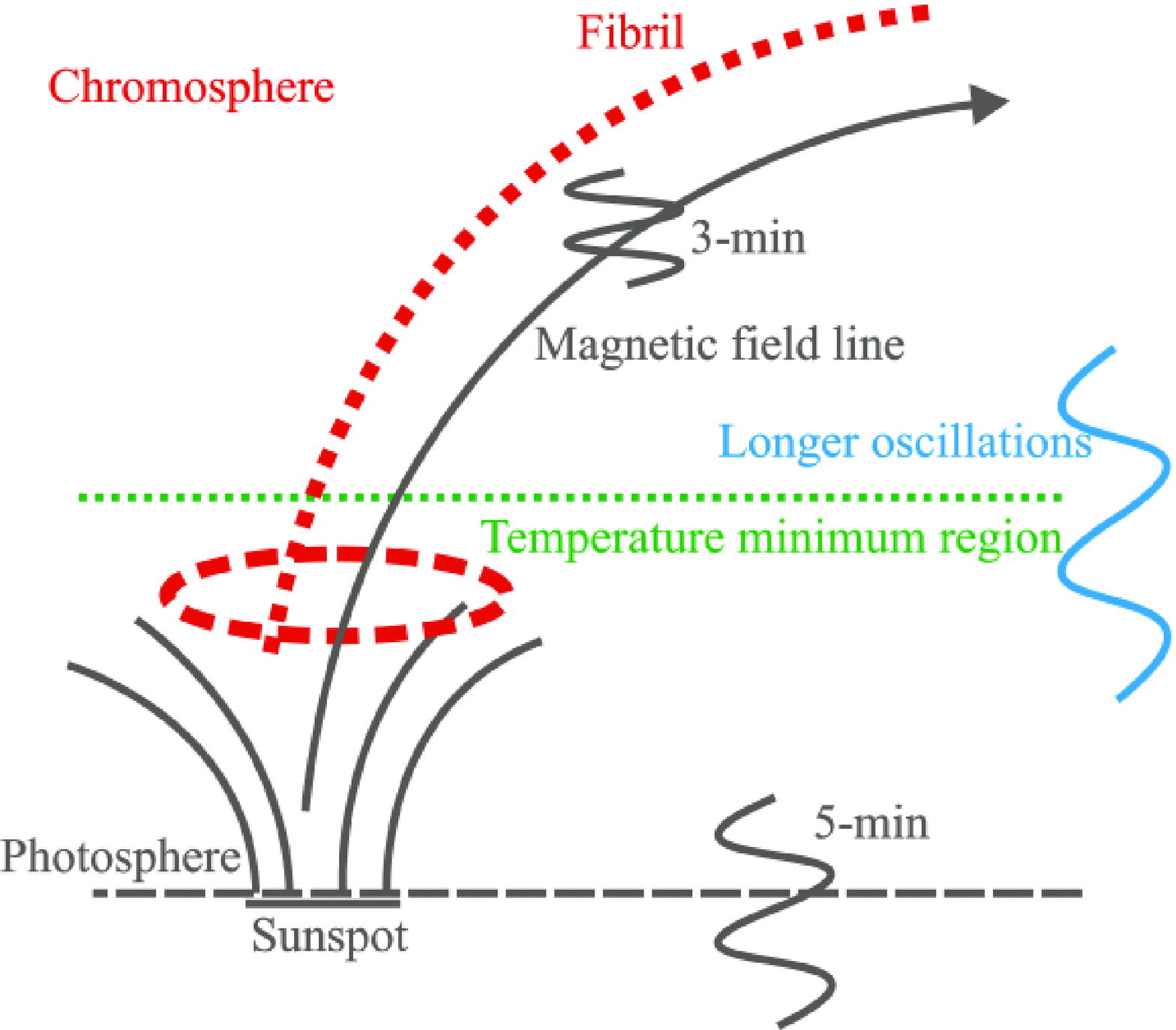}
             }
   \caption{The Illustration depicting the interaction between the chromosphere, photosphere and the role of fibril on connecting different layer.}
    \label{fibril_model_rev}
  \end{figure}

We propose a picture on how the photosphere related to chromosphere and how the fibril  play roles as the fabric of the interaction between the layers, as shown in Figure~\ref{fibril_model_rev}. As seen in Figure~\ref{fibril_model_rev}, even the dominating 5-minute, there were various frequencies occuring near the photosphere. In particular the long period of 15-minute in the blue wing of H$\alpha$, this area might play a role as the passing filter on frequency, as well as the magnetic-field interaction. It is very likely that in the same region, the fibril also formed due to filtering process just below the temperature minimum region. It explains how there is transparency in the temperature minimum region. At the same time, the fibril was also loading the material from below, which was brought by the magnetic field before it was uploaded below the chromosphere, (just above the red circle after departing from the temperature minimum region in Figure~\ref{fibril_model_rev}). We believe that it might be able to add to the picture of de la Cruz Rodr{\'{\i}}guez 
\& Socas-Navarro~(\cite{2011A&A...527L...8D}), that chromospheric fibrils are a visual proxy for the magnetic field lines and may need to be reconsidered. This does not necessarily mean that there might be different mechanism, but there might be filtering process to the magnetic field along the fibrilar pattern.

\subsection{Light Bridge Connection?}

From Figure~\ref{powermap}, 2$^{nd}$ column in H$\alpha$, both from \add{the} line center \& blue wing (3$^{rd}$ \& 2$^{nd}$ row), we observed a lane-like structure that looked like a light bridge (LB) structure. 

The structure can be identified from the average picture on G-band \& Ca II H (1$^{st}$ column, 4$^{th}$ \& 3$^{rd}$ row). Is the brightening in 15-minute oscillations also related to the LB? Louis et al.~(\cite{2008SoPh..252...43L}) suggested that LB could be the sites for heating the overlying chromosphere, based on the low-altitude reconnection, which can be seen as the brightening in the Ca images from Hinode observation.

We propose that if the LB plays a roles in the process, the mechanism would be that LB is the reconnection place (red ellipse in the Figure~\ref{fibril_model_rev}) which is brighter in 3-minute oscillations and the fibril is the ejected material due to reconnection. This argument has a basis, even there were no significant activities observed in the Ca II H \& G-band, however the surges in H$\alpha$ observed, with the LB  strengthened in the 3-minute oscillations, the 15-minute signifies the significance of the fibril.

Is this also related to umbral flash? The enhancement on the Doppler-signal in the 3-minute oscillations indicate the enhancement of the chromospheric line of H$\alpha$, but the dynamic in the fibril area does not necesarily define the whole dynamic of the umbrae.

\section{Conclusions}
\label{S-Conclusion}
We have shown the chromospheric dynamic and how it is related to the photosphere by studying the fibril  from high resolution in spatial and temporal observations. The filtering processes
in the photosphere region, particularly near the strong magnetic source and its inclination, also mass motion flow, added with short period oscillation and can trace the formation and process that connect between the layers. We have found that (1) the intermittent ejected material through fibril from Doppler images has clearly shown the oscillation, as seen in the H$\alpha$ blue wing and Doppler-signal images, (2) the oscillations in the umbrae and penumbrae magnetic field lines above the photosphere cause a gradual broadening and form the area like a ring shape from 3 to 15-minute oscillations as a function with height. These made a distinct boundary of umbrae and penumbrae which suggest the comb structure, and indicate that the oscillation could propagate along the inclined magnetic flux tubes from below, (3) the 3-minute oscillations strongly appeared in the inclined penumbrae magnetic filed lines and gave the clear boundary of the light bridge, as clearly seen in the H$\alpha$ line center and H$\alpha$ blue wing, (4) the 5-minute was dominated in the outer umbrae region near the boundary of umbrae-penumbrae, which suggest that the comb structure is where predominantly the downward motion, (5) the 7-minute was transparent in the H$\alpha$ blue wing. But the same with 10 and 15-minute, it concentrates in the inner penumbrae, as seen in the H$\alpha$ line center. But does the fibril also related to umbral flash? From the morphological signature it has indications of interplay between mass motion and strong magnetic field, but further analysis is still needed to reveal the interaction between the two.

\begin{acknowledgements}
Dutch Open Telescope, operated at the Spanish Observatorio del Roque de los Muchachos of the Instituto de Astrof\'{i}sica de Canarias.
\end{acknowledgements}

\label{lastpage}

\end{document}